\documentclass{article}
\usepackage{pdproc2} 

\usepackage{graphicx}
\usepackage{epsfig}
\begin{document}
\title{IceCube: The Rationale for Kilometer-Scale Neutrino Detectors}
\author{Francis Halzen}
\address{Department of Physics, University of Wisconsin, Madison, WI 53706,USA
\\and\\
DESY, Zeuthen, Germany}

\abstract{At a time when IceCube is nearing completion, we revisit the rationale for constructing kilometer-scale neutrino detectors.  We focus on the prospect that such observatories reveal the still-enigmatic sources of
  cosmic rays.  While only a ``smoking gun'' is missing for the
  case that the Galactic component of the cosmic-ray spectrum
  originates in supernova remnants, the origin of the extragalactic
  component remains a mystery. We speculate on neutrino emission from gamma-ray bursts and active galaxies.}

\section{The First Kilometer-Scale, High-Energy Neutrino Detector: IceCube}
\begin{figure}[h!]
    \begin{center}
\includegraphics[width=0.5\textwidth]{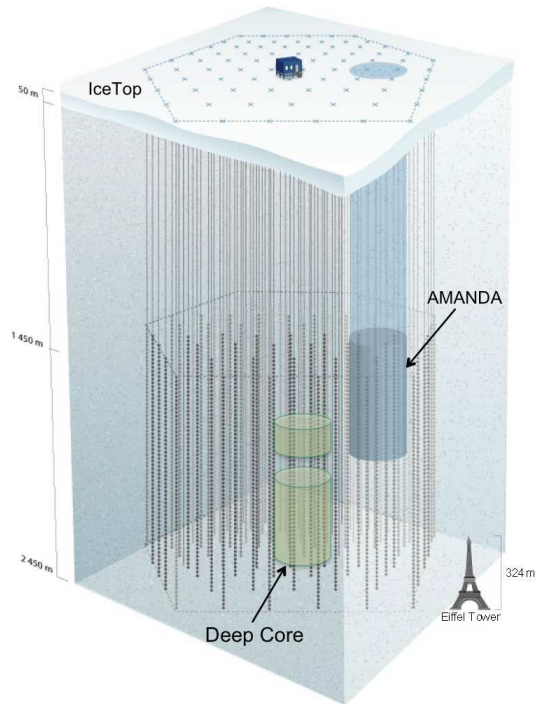}
    \end{center}
\caption {The IceCube detector, consisting of IceCube and IceTop 
and the low-energy sub-detector DeepCore. 
Also shown is the first-generation AMANDA detector.}
    \label{fig:deepcore}
 \end{figure}

\vspace{-0.25 cm}
A series of first-generation experiments\cite{Spiering:2008ux} have
demonstrated that high-energy neutrinos with $\sim10$\,GeV energy and
above can be detected by observing Cherenkov radiation from
secondary particles produced in neutrino interactions inside large
volumes of highly transparent ice or water instrumented with a lattice
of photomultiplier tubes. The first second-generation detector,
IceCube  (see Fig.~\ref{fig:deepcore}), is under construction at the geographic South
Pole\cite{Klein:2008px}. 

IceCube will consist of 80 km-length strings, each instrumented with 60 10-inch photomultipliers spaced 17\,m apart. The deepest module is located at a depth of 2450\,m so that
the instrument is shielded from the large background of cosmic rays at
the surface by approximately 1.5\,km of ice. Strings are arranged
at apexes of equilateral triangles that are 125\,m on a side. The
instrumented detector volume is a cubic kilometer of dark, highly
transparent and sterile Antarctic ice. The radioactive
background is dominated by instrumentation deployed into this
natural ice.

Each optical sensor consists of a glass sphere containing the
photomultiplier and the electronics board that digitizes the signals
locally using an on-board computer. The digitized signals are given a
global time stamp with residuals accurate to less than 3\,ns and are
subsequently transmitted to the surface. Processors at the surface
continuously collect these time-stamped signals from the optical
modules; each functions independently.  The digital messages are sent
to a string processor and a global event trigger. They are
subsequently sorted into the Cherenkov patterns emitted by secondary
muon tracks that reveal the direction of the parent
neutrino\cite{Halzen:2006mq}.

\begin{figure}
\begin{center}
\includegraphics[width=0.5\textwidth]{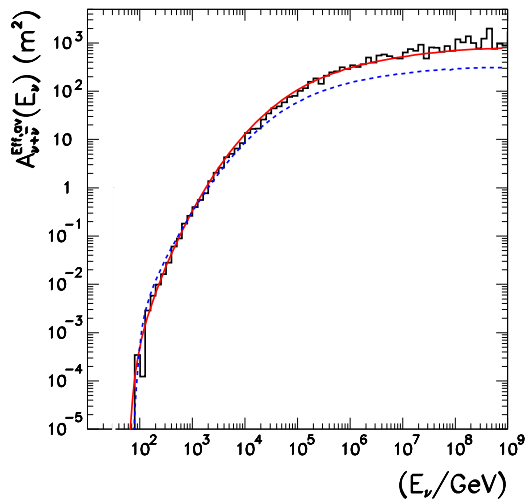}
\end{center}
\caption{The neutrino effective area (averaged over the Northern
Hemisphere) from IceCube simulation (black histogram) is compared to
the convolution of the approximate muon effective area from
reference\protect\cite{GonzalezGarcia:2009jc} (solid red line) that we will
use in the various estimates of event rates throughout this paper. The
neutrino area exceeds the design area (shown as the dashed blue line)\protect\cite{ic2004}
at high energy. }
\label{fig:areas}
\end{figure}

Based on data taken with 40 of the 59 strings that have already been deployed, the anticipated effective area of the completed IceCube detector is shown in Fig.~\ref{fig:areas}. Notice the factor 2 to 3 increase in effective area over what had been anticipated\cite{ic2004}.

Despite a discovery potential touching a wide range of scientific issues, construction of IceCube and KM3NeT\cite{Migneco:2008zz}, a concept for a similar detector in the Northern hemisphere, has been largely motivated by the possibility of opening a new window on the Universe using neutrinos as cosmic messengers. Specifically, we will revisit IceCube's prospects to detect cosmic neutrinos associated with cosmic rays and to reveal their sources prior to the 100th anniversary of their discovery by Victor Hess in 1912.

\begin{figure}
\begin{center}
\includegraphics[width=0.6\textwidth]{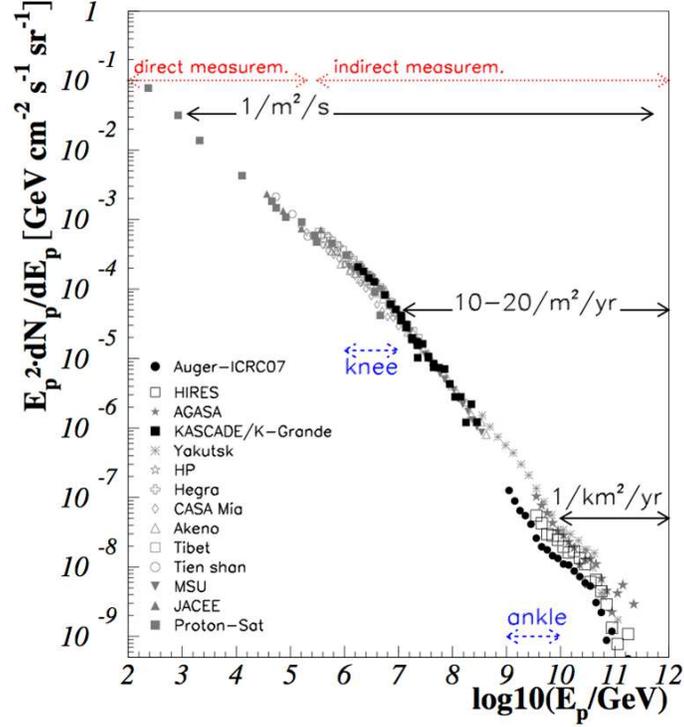}
\end{center}
\caption{At the energies of interest here, the cosmic-ray spectrum
follows a sequence of 3 power laws. The first 2 are separated by the
``knee'', the 2nd and 3rd by the ``ankle''. Cosmic
rays beyond the ankle are a new population of particles produced in
extragalactic sources.}
\label{fig:cr_spectrum}
\end{figure}

Cosmic accelerators produce particles with energies in excess of $10^8$\,TeV; we still do not know where or how\cite{Sommers:2008ji}. The flux of cosmic rays observed at Earth is shown in Fig.~\ref{fig:cr_spectrum}. The energy spectrum follows a sequence of three power laws. The first two are separated by a feature dubbed the ``knee'' at an energy\footnote{We will use energy units TeV, PeV and EeV, increasing by factors of 1000 from GeV energy.} of approximately 3\,PeV.  There is evidence that cosmic rays up to this energy are Galactic in origin.  Any association with our Galaxy disappears in the vicinity of a second feature in the spectrum referred to as the ``ankle"; see Fig.~\ref{fig:cr_spectrum}. Above the ankle, the gyroradius of a proton in the Galactic magnetic field exceeds the size of the Galaxy, and we are witnessing the onset of an extragalactic component in the spectrum that extends to energies beyond 100\,EeV. Direct support for this assumption now comes from two experiments\cite{Abraham:2008ru} that have observed the telltale structure in the cosmic-ray spectrum resulting from the absorption of the particle flux by the microwave background, the so-called Greissen-Zatsepin-Kuzmin cutoff. The origin of the flux in the intermediate region covering PeV-to-EeV energies remains a mystery, although it is routinely assumed that it results from some high-energy extension of the reach of Galactic accelerators.

Acceleration of protons (or nuclei) to TeV energy and above requires massive bulk flows of relativistic charged particles. These are likely to originate from exceptional gravitational forces in the vicinity of black holes or neutron stars. Gravity powers large currents of charged particles that are the origin of high magnetic fields. These create the opportunity for particle acceleration by shocks. It is a fact that electrons are accelerated to high energy near black holes; astronomers detect them indirectly by their synchrotron radiation. Some must accelerate protons, because we observe them as cosmic rays.

How many gamma rays and neutrinos are produced in association with the cosmic-ray beam?\footnote{We do not discuss cosmic neutrinos directly produced in the interactions of cosmic rays with microwave photons here; extensions of IceCube have been proposed to detect them\cite{Allison:2009rz}.} Generically, a cosmic-ray source should also be a beam dump. Cosmic rays accelerated in regions of high magnetic fields near black holes inevitably interact with radiation surrounding them, e.g., UV photons in active galaxies or MeV photons in gamma-ray--burst fireballs. In these interactions, neutral and charged pion secondaries are produced by the processes
\begin{equation}
p + \gamma \rightarrow \Delta^+ \rightarrow \pi^0 + p
\mbox{ \ and \ }
p + \gamma \rightarrow \Delta^+ \rightarrow \pi^+ + n.
\end{equation}

While secondary protons may remain trapped in the high magnetic fields, neutrons and the decay products of neutral and charged pions escape. The energy escaping the source is therefore distributed among cosmic rays, gamma rays and neutrinos produced by the decay of neutrons, neutral pions and charged pions, respectively. In the case of Galactic supernova shocks, cosmic rays interact with gas, e.g. with dense molecular clouds, as well as with radiation,
producing equal numbers of pions of all three charges in hadronic collisions $p+p \rightarrow n\,[\,\pi^{0}+\pi^{+} +\pi^{-}]+X$.

Kilometer-scale neutrino detectors have the sensitivity to reveal generic cosmic-ray sources with an energy density in neutrinos comparable to their energy density in cosmic rays\cite{TKG} and pionic TeV gamma rays\cite{AlvarezMuniz:2002tn}.

\section{Sources of Galactic Cosmic Rays}
\vspace{-0.25cm}

Supernova remnants were proposed as possible sources of Galactic cosmic rays as early as 1934 by Baade and Zwicky\cite{zwicky}; their proposal is still a matter of debate after more than 70 years.  Galactic cosmic rays reach energies of at least several PeV, the ``knee" in the spectrum. Their interactions with Galactic hydrogen in the vicinity of the accelerator should generate gamma rays from decay of secondary pions that reach energies of hundreds of TeV. Such sources should be identifiable by a relatively flat energy spectrum that extends to hundreds of TeV without attenuation; they have been dubbed PeVatrons. Straightforward energetics arguments are sufficient to conclude that present air Cherenkov telescopes should have the sensitivity necessary to detect TeV photons from PeVatrons\cite{GonzalezGarcia:2009jc}.

They may have been revealed by an all-sky survey in $\sim 10$\,TeV gamma rays with the Milagro detector\cite{Abdo:2006fq}. Sources have been identified that are located within nearby star-forming regions in Cygnus and in the vicinity of Galactic latitude $l=40$\,degrees; some are not readily associated with known supernova remnants or with non-thermal sources observed at other wavelengths.  Subsequently, directional air Cherenkov telescopes were pointed at three of the sources, identifying them as PeVatron candidates with an approximate $E^{-2}$ energy spectrum that extends to tens of TeV without evidence for a cutoff\cite{hesshotspot,magic2032}, in contrast with the best-studied supernova remnants RX J1713-3946 and RX J0852.0-4622 (Vela Junior). It remains to be seen of course whether any of the Milagro sources really do reach photon energies of hundreds of TeV.

Some Milagro sources may actually be molecular clouds illuminated by the cosmic-ray beam accelerated in young remnants located within $\sim100$\,pc. One expects indeed that multi-PeV cosmic rays are accelerated only over a short time period, when the remnant transitions from free expansion to the beginning of the Sedov phase and the shock velocity is high. The high-energy particles can produce photons and neutrinos over much longer periods when they diffuse through the interstellar medium to interact with nearby molecular clouds; for a detailed discussion, see reference 16. An association of molecular clouds and supernova remnants is expected in star-forming regions.

\begin{figure}
\centering
\includegraphics[width=3.0in]{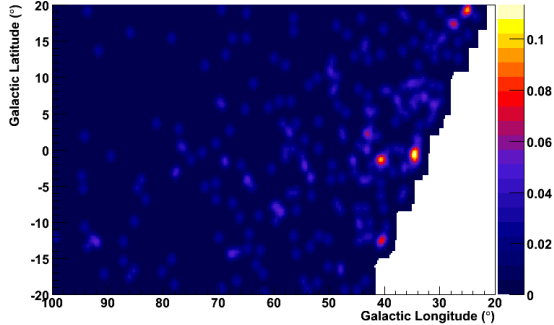}
\caption{Simulated sky map of IceCube in Galactic coordinates after 5
years of operation of the completed detector.  Two Milagro
sources are visible ``by eye" with 4 events for MGRO J1852+01 and 3
events for MGRO J1908+06 with energy in excess of 40\,TeV. These, as
well as the background events, have been randomly distributed according
to the resolution of the detector and the size of the sources.}
\label{fig:5year_Map_1}
\end{figure}

Despite the rapid development of instruments with improved sensitivity, it has been impossible to conclusively pinpoint supernova remnants as the sources of cosmic rays by identifying accompanying gamma rays of pion origin. Detecting the accompanying neutrinos would provide incontrovertible evidence for cosmic-ray acceleration in the sources. Particle physics dictates the relation between gamma rays and neutrinos and basically predicts the production of a $\nu_\mu+\bar\nu_\mu$ pair for every two gamma rays seen by Milagro. This calculation can be performed in a more sophisticated way with approximately the same outcome. We conclude that, within uncertainties in the source parameters and the detector performance, confirmation that Milagro mapped sources of Galactic cosmic rays should emerge after operating the complete IceCube detector for several years; see Fig.~\ref{fig:5year_Map_1}.

The quantitative statistics can be summarized as follows. For average values of the parameters describing the flux, we find that the completed IceCube detector could confirm sources in the Milagro sky map as sites of cosmic-ray acceleration at the
$3\sigma$ level in less than one year and at the $5\sigma$ level in three years\cite{GonzalezGarcia:2009jc}. These results agree with previous estimates\cite{hkm}. There are intrinsic ambiguities in this estimate. With the performance of IceCube now well understood, these are of astrophysical origin. In the absence of observation of TeV-energy supernova neutrinos by IceCube, the nature of sources that produce cosmic rays near the knee of the spectrum remains unresolved.

\section{Sources of the Extragalactic Cosmic Rays}
\vspace{-0.25cm}
Although there is no direct evidence that supernovae accelerate cosmic rays, the idea is generally accepted because of energetics: three supernovae per century converting a reasonable fraction of a solar mass into particle acceleration can accommodate the steady flux of cosmic rays in the Galaxy. Energetics also drive speculation on the origin of extragalactic cosmic rays.

By integrating the cosmic-ray spectrum in Fig.~\ref{fig:cr_spectrum} above the ankle\cite{TKG}, we find that the energy density of extragalactic cosmic rays in the Universe is $\rho_E \simeq 3 \times10^{-19}\rm\,erg\ cm^{-3}$. The power required for a population of sources to generate this energy density over the Hubble time of $10^{10}$\,years is $3 \times 10^{37}\rm\,erg\ s^{-1}$ per (Mpc)$^3$. The fireball producing a gamma-ray burst (GRB) converts a fraction of a solar mass into the acceleration of electrons, seen as synchrotron photons. The energy in extragalactic cosmic rays can be accommodated with the reasonable assumption that shocks in the expanding GRB fireball convert roughly the same energy into the acceleration of protons that are observed as cosmic rays\cite{waxmanbahcall}. It so happens that $2 \times 10^{51}$\,erg per GRB will yield the observed energy density in cosmic rays after $10^{10}$ years given that their rate is 300 per $\textrm{Gpc}^{3}$ per year. Therefore 300 GRBs per year over Hubble time produce the observed cosmic-ray energy density in the Universe, just as 3 supernovae per century accommodate the steady flux of cosmic rays in the Galaxy.

Cosmic rays and synchrotron photons coexist in the expanding GRB fireball prior to it reaching transparency and producing the observed GRB display. Their interactions produce charged and neutral pions
\begin{equation}
p + \gamma \rightarrow \Delta^+ \rightarrow \pi^0 + p
\mbox{ \ and \ }
p + \gamma \rightarrow \Delta^+ \rightarrow \pi^+ + n.
\end{equation}
with probabilities 2/3 and 1/3, respectively. Subsequently the pions decay into gamma rays and neutrinos that carry 1/2 and 1/4 of the energy of the parent pion. We here assume that the four leptons in the decay $\pi^{+} \rightarrow \nu_{\mu} + \left (e + \bar{\nu}_{e} + \nu_{\mu}\right)$ equally share the charged pion's energy. The energy of the pionic leptons relative to the proton is: 
\begin{equation}
x_{\nu} = \frac {E_{\nu}}{E_{p}} = \frac {1}{2} \langle 
x_{p \rightarrow \pi}\rangle \; \simeq \,\frac {1}{20},
\end{equation}
and
\begin{equation}
x_{\gamma} = \frac {E_{\gamma}}{E_{p}} = \frac {1}{2} 
\langle x_{p \rightarrow \pi}\rangle \;\simeq\, \frac {1}{10}.
\end{equation}
Here
\begin{equation}
\langle x_{p \rightarrow \pi}\rangle \, = \, \langle 
\frac{E_{\pi}}{E_{p}}\rangle \;\simeq \,0.2
\end{equation}
is the average energy transferred from the proton to the pion. The secondary neutrino and photon fluxes produced by a GRB are
\begin{eqnarray}
\displaystyle  \frac {dN_{\nu}}{dE} &=& 
\displaystyle 1\; \frac {1}{3}\; \frac {1}{x_{\nu}} \;\frac {dN_{p}}{dE_{p}}\left( \frac {E_{p}}{x_{\nu}}\right), \\
\displaystyle \frac {dN_{\gamma}}{dE} &=& 2
\displaystyle \; \frac {2}{3} \;\frac {1}{x_{\gamma}} \;\frac {dN_{p}}{dE_{p}} \left( \frac {E_{p}}{x_{\gamma}}\right) = \frac{1}{8}\; \frac {dN_{\nu}}{dE}. 
\end{eqnarray}
Here $N_{\nu} \left(=  N_{\nu_{\mu}} =  N_{\nu_{e}}=  N_{\nu_{\tau}}\right)$ represents the sum of the neutrino and antineutrino fluxes which are not distinguished by the experiments. Oscillations over cosmic baselines yield approximately equal fluxes for the 3 flavors.

Neutrinos reach us from sources distributed over all redshifts, while cosmic rays do so only from local sources inside the so-called GZK radius of less than 100\,Mpc. The evolution of the sources will boost the neutrino flux by a factor $f_{GZK}$ that depends on the poorly known redshift dependence of GRBs. Assuming a dependence not very different from galaxies, this factor is approximately 3, therefore
\begin{equation}
\frac {dN_{\nu}}{dE_{\nu}} = \frac {1}{3}\; x_{\nu}\; \frac {dN_{p}}{dE_{p}} \left ( \frac {E_{p}}{x_{\nu}} \right) \; f_{GZK} \simeq x_{\nu} \;\frac {dN_{p}}{dE_{p}} \left ( \frac {E_{p}}{x_{\nu}} \right).
\end{equation}
The rate of neutrinos actually detected from this flux can be approximately calculated in the usual way~\cite{GonzalezGarcia:2009jc}
\begin{equation}
N = 2\pi \times \rm area \times \rm time \times \int \frac {dN_{\nu}}{dE}\; P_{\nu \rightarrow \mu} \;dE,
\end{equation}
with
\begin{equation}
P_{\nu \rightarrow \mu} \simeq 10^{-6} \;E_{\nu}\left(\rm TeV\right),
\end{equation}
and
\begin{equation}
\frac {dN_{\nu}}{dE} \simeq x_{\nu} \; \frac {5 \times 10^{-11}}{E} \rm\,TeV \,cm^{-2}\, s^{-1}\, sr^{-1}.\end{equation}
Here we approximated the energy in the cosmic-ray flux to 1 particle per km\,$^2$ per year at 10\,EeV. We obtain
\begin{equation}
N  \simeq 4 \;log \left( \frac {E_{\nu_{max}}}{E_{\nu_{min}}} \right) \simeq 15\; \rm events \;\rm year^{-1}.
\end{equation}
The value of the log describing the energy reach of the accelerators is not important, as it cancels the same log that appears in the integration that yields $\rho_{E}$. The number of events is indeed 15 after you cancel the two logs. The calculation can be repeated for arbitrary power-law spectra $E^{-\left(\alpha+1\right)}$. The number of events is smaller and approaches 15 per kilometer square year when $\alpha$ approaches 1. This calculation neglects the partial absorption of neutrinos in the Earth, but this is compensated by the fact that the IceCube effective area exceeds $1\,\rm km^2$, as previously discussed. For details, see reference 19.

The key to the production of neutrinos is the fact that protons coexist with GRB gamma rays when the fireball is opaque to gamma-gamma and p-gamma interactions. The specific prediction is obtained because we assumed that the protons interacted once, i.e. one neutron and one charged and neutral pion are produced for every proton; this is referred to as a transparent source. In general,
\begin{equation}
\frac {dE_{\nu}}{dE_{\nu}} =  [ 1- \left(1-e^{-n_{int}}\right)] \; \frac {1}{3} \; x_{\nu} \; \frac {dN_{p}}{dE_{p}} \left ( \frac {E_{p}}{x_{\nu}} \right) \; f_{GZK} \simeq n_{int} \; x_{\nu} \; \frac {dN_{p}}{dE_{p}} \left ( \frac {E_{p}}{x_{\nu}} \right),
\end{equation}
where $n_{int}$ is the number of interactions of the proton before escaping the fireball; it is determined by the optical depth of the source for p\,$\gamma$ interactions.

The expanding GRB fireball is a shell of radius $R^{\prime}$ and width $\Delta RÕ^{\prime}$. The primes refer to the rest frame of the fireball which is boosted toward the observer by a factor $\Gamma$. The size of the GRB fireball is given by the light crossing speed $\Delta R^{\prime} = c \,\Delta t \,\Gamma$, where $\Delta t$ is the typical variability of the GRB emission. $\Delta t$ is assumed to be 10 milliseconds, on average. The boost factor $\Gamma$ between the expanding fireball and the observer is 300, on average. The number of $p\gamma$ interactions is given by

\begin{equation}
n_{int} =  \frac {\Delta R^{\prime}}{\lambda_{p\gamma}} = \left( c \, \Delta t \, \Gamma \right) \left( n^{\prime}_{\gamma} \,\sigma_{p\gamma} \right).
\end{equation}
The mean-free path of the photon is $\left( n^{\prime}_{\gamma} \sigma_{p\gamma} \right)^{-1}$, where the density of photons in the fireball $n^{\prime}_{\gamma}$  is obtained from the observed luminosity and the volume of the expanding fireball shell
\begin{equation}
n^{\prime}_{\gamma} = \frac {E_{tot}/E_{\gamma}}{V^{\prime}},
\end{equation}
with
\begin{equation}
V^{\prime} = 4 \pi R^{{\prime} 2} \Delta R^{\prime} = 4 \pi \left( c \,\Delta t \,\Gamma^{2} \right)^{2} \left( c \,\Delta t\, \Gamma \right).
\end{equation}
Here the total fireball energy is $E_{tot}$ is $2 \times 10^{51}$\,erg, and the average photon energy is $E_{\gamma} \simeq 1$\,MeV. From the above equations, one concludes indeed that $n_{int} \simeq 1$ as was previously assumed. The conventional choice that $\Delta t$ is 10\,milliseconds can be debated; for details, see reference 20.

If one alternatively assumes that protons are trapped in the source and neutrons escape to decay into the observed cosmic rays, then the observed flux has to be identified with the neutron flux given by
\begin{equation}
\frac {dN_{n}}{dE_{n}} = \frac {1}{3} \; \frac {1}{0.5} \; \frac {dN_{p}}{dE_{p}} \left( \frac {E_{p}}{0.5} \right),
\end{equation}
i.e. for every proton, one neutron is produced with half the energy. If $dN_{n}/dE_{n}$ represents the observed cosmic-ray flux, there is an increased proton flux inside the source to produce charged pions, and the neutrino flux is increased by a factor 6.

\begin{figure}
\centering
\includegraphics[width=0.5\textwidth]{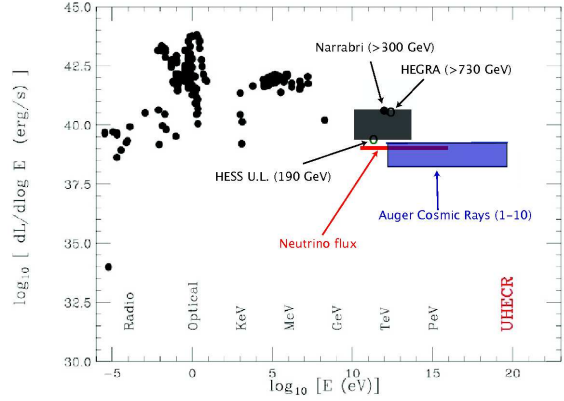}
\caption{Spectral energy distribution of Cen A (black dots). Keeping in
mind that the source is variable, we show our estimates for the flux
of TeV gamma rays (upper gray shading) and cosmic rays assuming that between
1 and 10 events observed by Auger originated at Cen A (lower blue
shading). We note that cosmic-ray and TeV gamma-ray fluxes
estimated in this paper are at a level of the electromagnetic
component shown from radio waves to GeV photons. Our estimate for the
neutrino flux is shown as the red line. }
\label{fig:CenA_SED_v4}
\end{figure}

Problem solved? Not really: the energy density of the extragalactic cosmic rays can be accommodated by active galactic nuclei (AGN) provided each converts  $2 \times 10^{44}\rm\,erg\ s^{-1}$ into particle acceleration. As is the case for GRBs, this is an amount that matches their output in electromagnetic radiation. The calculation just performed for GRBs can be repeated assuming that active galactic nuclei are the sources of the cosmic rays. In this case, the site of cosmic-ray production is a matter of speculation, and the nature of the target photons is uncertain. It is therefore difficult to estimate $n^{\prime}_{\gamma}$; for a discussion see reference 21. It has been argued on the basis of data~\cite{Halzen:2008vz,anchordoquicena} that Cen A and M87 may emit similar energy in TeV photons and cosmic rays, suggesting that  $n_{int} \simeq 1$; see Fig.~\ref{fig:CenA_SED_v4}.  The interaction length of the protons cannot be much larger; otherwise, the TeV photons would be absorbed in the source by gamma-gamma interactions. This is the Waxman-Bahcall bound~\cite{WB} which follows from the fact that the $\gamma\gamma$ cross section in the TeV region happens to be equal to the $p\gamma$ cross section in the EeV region.

Whether GRB or AGN, the observation that these sources radiate similar energies in photons and cosmic rays is consistent with the beam dump scenario previously discussed. In the interaction of cosmic rays with radiation and gases near the black hole, roughly equal energy goes into the secondary protons (or neutrons) and neutral pions whose energy ends up in cosmic rays and gamma rays, respectively. It predicts a matching flux of neutrinos that, after many correction factors that cancel, is roughly equal to, or a fraction of, the cosmic-ray flux previously introduced:
\begin{equation}
E_{\nu}^{2} dN / dE_{\nu}= 5 \times 10^{-11}\rm\,TeV \,cm^{-2}\, s^{-1}\, sr^{-1}
\label{extragalactic}
\end{equation}

\begin{figure}[!h]
\centering
\includegraphics[width=2.75in, angle=270]{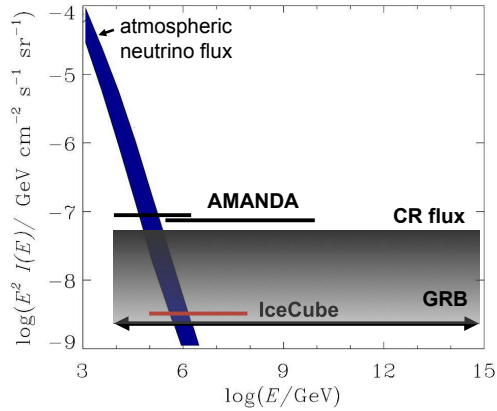}
\caption{Our energetics estimate of the flux of neutrinos associated
with sources of the highest-energy cosmic rays (the shaded region in the plot ranging from the cosmic-ray flux to the flux derived assuming that GRBs are the sources of extragalactic cosmic rays) is compared to upper limits established by AMANDA and the sensitivity of IceCube\,\protect\cite{merida}. Also shown is the background flux of atmospheric neutrinos at lower energies.}
\label{fig:agn_spectrum}
\end{figure}

If we adjust it downward by a factor $x_{\nu}$, we obtain a generic neutrino flux predicted by the GRB and AGN scenarios. After seven years of operation, AMANDA's sensitivity is approaching the interesting range, but it takes IceCube to explore it; see Fig.~\ref{fig:agn_spectrum}.

If it turns out that GRBs are the sources, IceCube's mission is relatively straightforward, because we expect to observe of order 10 neutrinos per kilometer square per year~\cite{guetta} in coincidence with Swift and Fermi GRBs, which translates into a $5\sigma$ observation.  A similar statistical power can be obtained by detecting showers produced by electron and tau neutrinos.

In summary, while the road to identification of the sources of the Galactic cosmic ray has been mapped, the origin of the extragalactic component remains as mysterious as ever.

A more detailed version of this talk can be found in the proceedings of the  ``XIII International Workshop on Neutrino Telescopes",  ``Istituto Veneto di Scienze, Lettere ed Arti", Venice, Italy, edited by M. Baldo Ceolin, p. 17 (2009).

\section{Acknowledgments} 
\vspace{-0.25cm}
I would like to thank my collaborators Elisa Bernardini, Concha Gonzalez-Garcia, Darren Grant, Alexander Kappes and Aongus O'Murchadha, as well as John Beacom, Julia Becker, Peter Biermann, Steen Hannestad, Christian Spiering and Stefan Westerhoff for valuable discussions. This research was supported in part by the U.S. National Science Foundation under Grants No.~OPP-0236449 and  PHY-0354776; by the University of Wisconsin Research Committee with funds granted by the Wisconsin Alumni Research Foundation; and by the Alexander von Humboldt Foundation in Germany.

\end{document}